\documentstyle[12pt]{article}

\input{tcilatex}

\begin{document}

\title{Hamiltonian and measuring time for analog quantum search}
\author{Jin-Yuan Hsieh$^{1}$, Che-Ming Li$^{2}$, and Der-San Chuu$^{2}$ \\
$^{1}$Department of Mechanical Engineering, Ming Hsin University \\
of Science and Technology, Hsinchu 30441,Taiwan.\\
$^{2}$Institute and Department of Electrophysics, National Chiao\\
Tung University, Hsinchu 30050, Taiwan.}
\maketitle

\begin{abstract}
We derive in this study a Hamiltonian to solve with certainty the analog
quantum search problem analogue to the Grover algorithm. The general form of
the initial state is considered. Since the evaluation of the measuring time
for finding the marked state by probability of unity is crucially important
in the problem, especially when the Bohr frequency is high, we then give the
exact formula as a function of all given parameters for the measuring time.
\end{abstract}

To solve a search problem, the remarkable Grover quantum algorithm\cite%
{grover1} provides a quadratic speedup over its classical counterpart. If
there are $M$ items among $N$ unsorted items to be found, then using the
Grover algorithm will accomplish the computation in $O(\sqrt{N})$ quantum
mechanical steps, instead of $O(N)$ classical steps. Zalka\cite{zalka}
further has shown that Grover's algorithm is optimal since it needs minimal
oracle calls to do this job. The Grover algorithm is carried out by a
successive iterations of operation on an initial state, which is usually
prepared by a uniform superposition of all states, to amplify the amplitude
of the marked state which is an uniform superposition of all target items.
Each iteration of the Grover algorithm is composed of the unitary
transformations of the $\pi $-inversion of the marked state and the $\pi $%
-inversion about average. We can think of that in a two-dimensional Hilbert
space the Grover algorithm in each step rotates a state to another by a
constatnt, finite angle. It then is rational to understand that the object
can never be reached exactly unless the target items are $N/4$ among $N$
items. In orther words, the marked states can only be approached
asymptotically as $N$ is large, since then the rotating angle in each step
is small. Neverthless, it has been proposed that no matter whether $N$ is
large or not, modified phase-rotations can be applied in replacement of the $%
\pi $-inversions in the algorithm to search the marked state with certainty%
\cite{hsieh and li}. As a matter of fact, the Grover algorithm belongs to
the quantum cricuit model since it is carried out conventionally by a
sequence of universal quantum logic gates.

On the other hand, several researchers have proposed another way to solve
the quantum search problem. It is proposed that the quantum search
computation can be accomplished by controlled Hamiltonian time evolution of
a system, obeying the Schr\"{o}dinger equation

\begin{equation}
i\frac{d\left| \Psi (t)\right\rangle }{dt}=H(t)\left| \Psi (t)\right\rangle ,
\end{equation}%
where $\hbar =1$ is imposed for convenience. Farhi and Gutmann\cite{Farhi
and Gutmann} first presented the Hamiltonian $H_{fg}=E_{fg}(\left|
w\right\rangle \left\langle w\right| +\left| s\right\rangle \left\langle
s\right| )$, where $\left| w\right\rangle $ is the marked state and $\left|
s\right\rangle $ denotes the initial state. After the presentation of $%
H_{fg} $, Fenner\cite{Fenner} proposed another Hamiltonian $%
H_{f}=E_{f}i(\left| w\right\rangle \left\langle s\right| -\left|
s\right\rangle \left\langle w\right| )$. Recently, Bae and Kwon\cite{Bae and
Kwon} further proposed a generalized quantum search Hamiltonian

\begin{equation}
H_{g}=E_{fg}(\left| w\right\rangle \left\langle w\right| +\left|
s\right\rangle \left\langle s\right| )+E_{f}(e^{i\phi }\left| w\right\rangle
\left\langle s\right| +e^{-i\phi }\left| s\right\rangle \left\langle
w\right| )\text{.}
\end{equation}%
where $\phi $ is an additional phase to the Fenner Hamiltonian. Unlike the
Grover algorithm, which operates on a state in discrete time, a search
Hamiltonian \ operates a state in continuous time, so the $100\%$
probability for finding the marked state can be guaranteed. Both the
Hamiltonians $H_{fg}$ and $H_{f}$ can help to find the marked state with $%
100\%$ success. Bae and Kwon addressed that the generalized Hamiltonian $%
H_{g}$ can accomplish the search with certainty only when $\phi =n\pi $ is
imposed, where $n$ is arbitrary integer. In this work, however, we will show
that $H_{g}$ is in fact a Hamiltonian for the quantum search with certainty
and $\phi $ can be chosen arbitrarily to influence the required measuring
time in the task. Since Hamiltonian is considered, the energy-time relation
then will play an essential role in the problem. The evaluation of the
measuring time for the quantum search with certainty therefore becomes
crucially important. In this study, we then intend to derive the general
Hamiltonian for the time-controlled quantum search system first. Then the
exact time for measuring \ the marked state by probability of unity as a
function of all given conditions will be deduced.

Suppose that a two-dimensional, complex Hilbert space is spanned by the
orthonormal set of basis states $\left| w\right\rangle $, which is the
marked state, and $\left| w_{\bot }\right\rangle $. An initial state $\left|
s\right\rangle =\left| \Psi (0)\right\rangle $ is designed to evolve under a
time-independent quantum search Hamiltonian given by

\begin{equation}
H=E_{1}\left| E_{1}\right\rangle \left\langle E_{1}\right| +E_{2}\left|
E_{2}\right\rangle \left\langle E_{2}\right| \text{,}
\end{equation}%
where $E_{1}$ and $E_{2}$ are two eigenenergies of the quantum systum, $%
E_{1}>E_{2}$, and $\left| E_{1}\right\rangle $ and $\left|
E_{2}\right\rangle $ are the corresponding eigenstates satisfying the
completeness condition $\left| E_{1}\right\rangle \left\langle E_{1}\right|
+\left| E_{2}\right\rangle \left\langle E_{2}\right| =1$. In general, since $%
\left\langle E_{1}|E_{2}\right\rangle =0$, the eigenstates can be assumed by

\begin{equation}
\left| E_{1}\right\rangle =e^{i\alpha }\cos (x)\left| w\right\rangle +\sin
(x)\left| w_{\bot }\right\rangle \text{, and }\left| E_{2}\right\rangle
=-\sin (x)\left| w\right\rangle +e^{-i\alpha }\cos (x)\left| w_{\bot
}\right\rangle \text{.}  \nonumber
\end{equation}%
where $x$ is now an unknown constant and will be determined later due to the
required maximal probability for measuring the marked state. By the
assumptions given in (4), the Hamiltonian then can be written in the matrix
form

\begin{equation}
H=\left[ 
\begin{array}{cc}
E_{p}+E_{o}\cos (2x) & E_{o}\text{sin}(2x)e^{i\alpha } \\ 
E_{o}\text{sin}(2x)e^{-i\alpha } & E_{p}-E_{o}\cos (2x)%
\end{array}%
\right] \text{.}
\end{equation}%
where $E_{p}=(E_{1}+E_{2})/2$ and $E_{o}=(E_{1}-E_{2})/2$. The major
advantage of using the controlled Hamiltonian time evolution is that the
marked state $\left| w\right\rangle $ can always be searched with certainty.
The crucial key of the present problem in turn is to decide when to measure
the marked state by the probability of unity. So in what follows we will in
detail deduce the relation between the unknown $x$ and all the given
conditions and evaluate the exact measuring time for finding the marked
state with certainty.

The time evolution of the initial state, according to the Schr\"{o}dinger
equation (1), is given by $\left| \Psi (t)\right\rangle =e^{-iHt}\left|
s\right\rangle $. We wish to find the marked state with certainty, so the
probability of unity for finding the marked state, $P=\left| \left\langle
w\right| e^{-iHt}\left| s\right\rangle \right| ^{2}=1-\left| \left\langle
w_{\bot }\right| e^{-iHt}\left| s\right\rangle \right| ^{2}=1$, is required.
The general form of the initial state considered in this study is given by

\begin{equation}
\left| s\right\rangle =e^{iu}\sin (\beta )\left| w\right\rangle +\cos (\beta
)\left| w_{\bot }\right\rangle
\end{equation}%
where a nonzero phase $u$ may arise due to phase decoherence. According to
the expression $e^{-iHt}=e^{-iE_{1}t}\left| E_{1}\right\rangle \left\langle
E_{1}\right| +e^{-iE_{2}t}\left| E_{2}\right\rangle \left\langle
E_{2}\right| $, we therefore have

\begin{eqnarray}
\left\langle w_{\bot }\right| e^{-iHt}\left| s\right\rangle
&=&e^{-iE_{p}t}((\cos (\beta )\cos (E_{o}t)-\sin (\alpha -u)\sin (2x)\sin
(\beta )\sin (E_{o}t))  \nonumber \\
&&+i(\cos (2x)\cos (\beta )-\cos (\alpha -u)\sin (2x)\sin (\beta ))\sin
(E_{o}t))\text{.}
\end{eqnarray}

To accomplish the quantum search with certainty, $\left\langle w_{\bot
}\right| e^{-iHt}\left| s\right\rangle =0$ is required, so the
time-independent term $(\cos (2x)\cos (\beta )-\cos (\alpha -u)\sin (2x)\sin
(\beta ))$ in (7) must vanish and we thus can determine the unknown $x$ by

\begin{equation}
\cos (2x)=\frac{\sin (\beta )\cos (\alpha -u)}{\cos (\gamma )}\text{, or }%
\sin (2x)=\frac{\cos (\beta )}{\cos (\gamma )}\text{,}
\end{equation}%
where $\gamma $ is defined by $\sin (\gamma )=\sin (\beta )\sin (\alpha -u)$%
. As $x$ is determined by (8), the probability then becomes

\begin{eqnarray}
P &=&1-\left| \left\langle w_{\bot }\right| e^{-iHt}\left| s\right\rangle
\right| ^{2}  \nonumber \\
&=&1-\frac{\cos ^{2}(\beta )}{\cos ^{2}(\gamma )}\cos ^{2}(E_{0}t+\gamma ) 
\nonumber \\
&\geq &1-\frac{\cos ^{2}(\beta )}{\cos ^{2}(\gamma )}\text{.}
\end{eqnarray}%
Expression (9) obviously indicates that, by letting $\cos ^{2}(E_{0}t+\gamma
)=0$, we can measure the marked state by the probability of unity at the
time instants

\begin{equation}
t_{j}=\frac{(2j-1)\pi /2-\sin ^{-1}(\sin (\beta )\sin (\alpha -u))}{E_{o}}%
\text{, }j=1,2,...\text{.}
\end{equation}%
In what follows, let us focus on the first instant $t_{1}=(\pi /2-\sin
^{-1}(\sin (\beta )\sin (\alpha -u)))/E_{o}$. It is clear that a larger $%
E_{o}$ will lead to a shorter time to measure the marked state with
certainty. As can be seen in (9), the probability for measuring the marked
state, however, varies with time as a periodic function whose frequency is
the Bohr frequency $E_{o}/\pi $, so a larger $E_{o}$, on the contrast, will
also result in a more difficult control on the measuring time. That is, the
measuring time should be controlled more precisely as the Bohr frequency is
higher since then a small error in the measuring time will cost a serious
drop of the probability.

By the relations shown in (8), the present Hamiltonian now becomes

\begin{equation}
H=\left[ 
\begin{array}{cc}
E_{p}+E_{o}\frac{\sin (\beta )\cos (\alpha -u)}{\cos (\gamma )} & E_{o}\frac{%
\cos (\beta )}{\cos (\gamma )}e^{i\alpha } \\ 
E_{o}\frac{\cos (\beta )}{\cos (\gamma )}e^{-i\alpha } & E_{p}-E_{o}\frac{%
\sin (\beta )\cos (\alpha -u)}{\cos (\gamma )}%
\end{array}%
\right] \text{,}
\end{equation}%
which is represented by the energies $E_{p}$, $E_{o}$, and the phase $\alpha 
$. Alternatively, if we let

\begin{eqnarray}
E_{fg} &=&(E_{p}-E_{o}\frac{\sin (\beta )\cos (\alpha -u)}{\cos (\gamma )}%
)/\cos ^{2}(\beta )\text{,}  \nonumber \\
\text{ }E_{f}e^{i(\phi -u)} &=&\frac{E_{o}}{\cos (\gamma )}e^{i(\alpha
-u)}-E_{fg}\sin (\beta )\text{,}
\end{eqnarray}%
or inversely,

\begin{eqnarray}
E_{p} &=&E_{fg}+E_{f}\cos (\phi -u)\sin (\beta )\text{,}  \nonumber \\
\text{ }E_{o} &=&((E_{f}\cos (\phi -u)+E_{fg}\sin (\beta
))^{2}+E_{f}^{2}\sin ^{2}(\phi -u)\cos ^{2}(\beta ))^{\frac{1}{2}}\text{,}
\end{eqnarray}%
then the Hamiltonian can also be expressed by

\begin{equation}
H=\left[ 
\begin{array}{cc}
E_{fg}(1+\sin ^{2}(\beta ))+2E_{f}\cos (\phi -u)\sin (\beta ) & 
e^{iu}(E_{f}e^{i(\phi -u)}+E_{fg}\sin (\beta ))\cos (\beta ) \\ 
e^{-iu}(E_{f}e^{-i(\phi -u)}+E_{fg}\sin (\beta ))\cos (\beta ) & E_{fg}\cos
^{2}(\beta )%
\end{array}%
\right] \text{,}
\end{equation}%
which is represented by the energies $E_{fg}$ and $E_{f}$ and the phase $%
\phi $. Note that the expression of the present Hamiltonian (14) in fact is
equivalent to $H_{g}$ shown in (2), where the phase $u$, embedded in $\left|
s\right\rangle $, does not appear. Corresponding to both the presentations
(11) and (14), the exact first measuring time for finding the marked state $%
\left| w\right\rangle =\left| \Psi (t_{1})\right\rangle $ with $100\%$
success is at

\begin{eqnarray}
t_{1} &=&\frac{\frac{\pi }{2}-\sin ^{-1}(\sin (\beta )\sin (\alpha -u))}{%
E_{o}}  \nonumber \\
&=&\frac{\frac{\pi }{2}-\sin ^{-1}(\frac{E_{f}\sin (\beta )\sin (\phi -u)}{%
((E_{f}\cos (\phi -u)+E_{fg}\sin (\beta ))^{2}+E_{f}^{2}\sin ^{2}(\phi -u))^{%
\frac{1}{2}}})}{((E_{f}\cos (\phi -u)+E_{fg}\sin (\beta ))^{2}+E_{f}^{2}\sin
^{2}(\phi -u)\cos ^{2}(\beta ))^{\frac{1}{2}}}\text{.}
\end{eqnarray}

For the usual case $u=0$, if $\phi =n\pi $ or $\alpha =n\pi $ is impored,
then expression (14) reduces to the Hamiltonian considered by Bae and Kwon%
\cite{Bae and Kwon} to serve for a search with certainty, viz.,

\begin{eqnarray}
H_{p} &=&\left[ 
\begin{array}{cc}
E_{p}\pm E_{0}\cos (\beta ) & \pm E_{0}\cos (\beta ) \\ 
\pm E_{0}\cos (\beta ) & E_{p}\mp E_{0}\cos (\beta )%
\end{array}%
\right]   \nonumber \\
&=&\left[ 
\begin{array}{cc}
E_{fg}(1+\sin ^{2}(\beta ))\pm 2E_{f}\sin (\beta ) & (E_{fg}\sin (\beta )\pm
E_{f})\cos (\beta ) \\ 
(E_{fg}\sin (\beta )\pm E_{f})\cos (\beta ) & E_{fg}\cos ^{2}(\beta )%
\end{array}%
\right] \text{,}
\end{eqnarray}%
where $E_{0}=E_{f}\pm E_{fg}\sin (\beta )$ and $E_{p}=E_{fg}\pm E_{f}\sin
(\beta )$ are deduced. Bae and Kwon concluded that the generalized
Hamiltonian $H_{g}$ shown by (2), or $H$ given in (14) for $u=0$, can
provide the $100\%$ success in finding the marked state only when reducing
to $H_{p}$ if $\phi =n\pi $. We have shown that, however, the Hamiltonian $%
H_{g}$ can in fact accomplish a quantum search with certainty no matter what
value the phase $\phi $ is chosen to be. As shown in (15), the phase $\phi $
can be chosen arbitrarily to vary the measuring time. Also for $u=0$, the
present Hamiltonian obviously reduces to the Farhi and Gutmann Hamiltonian
if $E_{f}=0$, or inversely, if $E_{0}=E_{p}\sin (\beta )$ and $\alpha =0$ is
imposed and to the Fenner Hamiltonian when $E_{fg}=0$ and $\phi =\pi /2$, or
when $E_{p}=0$ and $\alpha =\pi /2$, is chosen. In the former case $%
E_{fg}=E_{p}$ is deduced while in the latter, $E_{o}=E_{f}\cos (\beta )$.

To summerize, we have derived in this study the quantum search Hamiltonian
in which the general initial state is considered. Given the initial
conditions for $\beta $ and $u$, this Hamiltonian can be either represented
by the energies $E_{p}=(E_{1}+E_{2})/2$, $E_{o}=(E_{1}-E_{2})/2$, and the
phase $\alpha $, or by the energies $E_{fg}$ and $E_{f}$ \ and the phase $%
\phi $, as shown in expressions (11) and (14), respectively. The shortest
time required to measure the marked state by probability of unity is
crucially important in the problem considered in this study, so it has also
been deduced in an exact formula as shown by expression (15).

J. Y. H. is supported in part by the National Science Council under the
grant No. NSC 90-2212-E-159-007.

\end{document}